\documentclass[aps,prl,reprint,superscriptaddress,showpacs]{revtex4-1}
\usepackage{graphicx}
\usepackage{amsmath}
\usepackage{amssymb}
\usepackage{multirow}
\usepackage{color}
\usepackage[normalem]{ulem}
\usepackage{comment}
\usepackage{dsfont}

\newcommand{\mlcancel}[1]{{\iffalse #1 \fi}}

\begin{document}


\title{Synchronized Switching in a Josephson Junction Crystal}


\author{Martin Leib}
\email[]{MartinLeib@circuitqed.net}
\affiliation{Technische Universit\"at M\"unchen, Physik Department, James Franck Str., 85748 Garching, Germany}
\author{Michael J. Hartmann}
\email[]{m.j.hartmann@hw.ac.uk}
\affiliation{Institute of Photonics and Quantum Sciences, Heriot-Watt University, Edinburgh, EH14 4AS, United Kingdom.}
\affiliation{Technische Universit\"at M\"unchen, Physik Department, James Franck Str., 85748 Garching, Germany}

\date{\today}

\begin{abstract}
We consider a superconducting coplanar waveguide resonator where the central conductor is interrupted by a series of uniformly spaced Josephson junctions. 
The device forms an extended medium that is optically nonlinear on the single photon level with normal modes that inherit the full nonlinearity of the junctions but are nonetheless accessible via the resonator ports. 
For specific plasma frequencies of the junctions a set of normal modes clusters in a narrow band and eventually become entirely degenerate.
Upon increasing the intensity of a red detuned drive on these modes, we observe a sharp and synchronized switching from low occupation quantum states to high occupation classical fields, accompanied by a pronounced jump from low to high output intensity.  
\end{abstract}


\maketitle
Achieving strong optical nonlinearities or appreciable effective interactions between individual photons is a long standing goal of Quantum Optics. As their generation requires strong coupling of photons to a nonlinear medium,
spatially very localized nonlinearities for light fields that are confined to the very small volumes of micro-cavities have successfully been realized
\cite{Birnbaum05,Hennessy07}. In recent years, the objective has thus moved on towards realizing strongly nonlinear optical response in
multiple, coupled cavities  \cite{LPOR:LPOR200810046,Leib2010,1367-2630-14-7-075024,Leib2013,Jin2013} or extended volumes doped with nonlinear media.
The description of light fields propagating in such devices can no longer invoke the classical or semiclassical approximations used in linear or nonlinear optics and thus forms a novel paradigm.
Experimentally, achieving appreciable photon-photon interactions despite the lower field amplitudes in larger volumes is a main challenge,
where Rydberg atoms with their strong dipole-dipole interaction \cite{Peyronel12,Baur13,Maxwell13} are one possible candidate.  

Here, we consider a long waveguide with closed ends that forms a strongly elongated, one-dimensional cavity which couples to many nonlinear scatterers. Yet, despite the large longitudinal extension, the light-matter coupling is ultra-strong, i.e. the vacuum Rabi frequency is comparable to the photon frequency, and the light fields inherit the full nonlinearity of the scatterers. Such ultra-strong coupling can be nicely reached in a novel discipline which bridges the gap between quantum optics and solid state physics, circuit quantum electrodynamics (cQED).

cQED setups have been used to simulate quantum optical phenomena \cite{Wallraff2004,Astafiev2007}. Lately they strive to conquer regimes that are elusive to photonic experiments at optical frequencies \cite{Egger13}, either with ultrastrong coupling \cite{Niemczyk2010} or unprecedented precision in deterministic steering of quantum mechanical states \cite{Mariantoni11,Lucero2012}. Here, we consider a long superconducting coplanar waveguide resonator (CPWR) in which ultra-strong coupling to Josephson junctions (JJs) is reached by integrating  the JJs directly into the CPWR's central conductor \cite{Castellanos08,Jung13}. The distances between the uniformly spaced JJs are chosen to be comparable to the wavelengths of the microwave photons we consider. Therefore the individual pieces of coplanar waveguide between adjacent JJs behave like individual CPWRs themselves thus forming a situation similar to electrons moving in the periodic potential of a crystal \cite{Zueco2012}. This spacing between the individual JJs promises to take the high precision and tunability of cQED setups into the realm of Josephson junction arrays \cite{Fazio2001}.

The extraordinary high coupling strength between the JJs and the CPWR demand for new ways of modeling the setup. Instead of devising an effective model of CPWR and JJs individually and then taking their coupling into account to obtain a Dicke model \cite{Dicke1954}, we consider the CPWR with JJs to be an indivisible entity and solve exactly for the eigenmodes of the harmonic part of the system in the manner of black box circuit quantization \cite{Nigg2012}. We then introduce the nonlinearity of the JJs as a perturbation to the linear dynamics of the eigenmodes. In this way we avoid the conceptual difficulties of virtual excitations in the ground state and the related complications in deriving the correct dissipative behavior \cite{Ridolfo2012}. 

For exploring the propagation of photons through our device we consider the CPWR to be continuously driven by a coherent input from one side and investigate its output at the opposite end.
In doing so we focus our attention on specific values of the plasma frequency of the JJs, where a set of modes clusters in a very narrow frequency band
and eventually becomes degenerate. It is thus also a candidate for a diabolical point \cite{Berry84}. We find that as the strength of a red detuned pump is increased, the modes synchronously switch from a quantum regime where they respond with a phase delay of $\pi$ to the drive (as two level systems do) to a classical regime where they are in phase with the drive and accumulate substantial amplitude.
The transition becomes increasingly sharper and more synchronized between the modes the closer the plasma frequency comes to the degeneracy point. 
As an observable signature of this phenomenon the output intensity jumps from a very low to a high value as the quantum to classical threshold is crossed by the drive amplitude.
The switching phenomenon we observe here thus features aspects of a quantum to classical transition for the cavity fields and a tendency of coupled JJs 
to synchronize their phases, which has been studied in their classical nonlinear dynamics \cite{Watanabe1994,Strogatz2004}.

\paragraph{Model\label{model}}
 \begin{figure}
\includegraphics[width=\columnwidth]{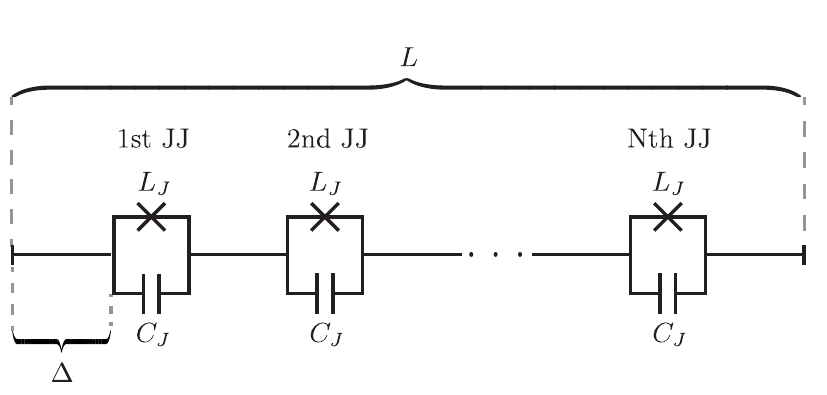}
\caption{Central conductor of a CPWR of length $L$ interrupted by $N$ identical and uniformly  distributed Josephson Junctions with Josephson inductance $L_J$ and capacitance $C_J$. \label{CPWresonator}}
\end{figure}
We consider a CPWR of length $L$ that supports one-dimensional current-charge waves with phase-velocity $v=1/\sqrt{lc}$ and
wave-impedance $Z_{0}=\sqrt{l/c}$, where $l$ and $c$ are its inductance and capacitance per unit length. The central conductor of the CPWR is interrupted by $N$ identical and uniformly distributed JJs with Josephson inductance $L_J$ and capacitance $C_J$. The plasma frequency $\omega_p=1/\sqrt{L_JC_J}$ of the JJs may be tuned either in advance via the design or in situ by using split ring dc-superconducting quantum interference devices (dc-SQUIDs) and threading a static bias flux through their rings.  The resonator is terminated on both sides open-circuited, enforcing current nodes at both ends. The state of the CPWR and the JJs can be described by the flux function $\phi(x,t)=\int_{-\infty}^t V(x,t')dt'$, with $V(x,t)$ the electrical potential of the CPWR at point $x$ with respect to the surrounding ground-plane. Physical observables like the excess charge per unit length, $Q=c\dot{\phi}$, or the current, $I=(\partial_x\phi)/l$, may be derived from $\phi$. The Lagrangian of the whole setup reads,
\begin{equation} \label{eq:;lagrangian}
\mathcal{L}=\sum_{j=1}^{N+1}\mathcal{L}_{j}^{\text{CPW}}+\sum_{j=1}^N\mathcal{L}_{j}^{\text{JJ}}\,,
\end{equation}
where,
$\mathcal{L}^{\text{CPW}}_j=\int_{(j-1)\Delta}^{j\Delta}\left\{\frac{c}{2}[\partial_t \phi(x,t)]^2-\frac{1}{2l}[\partial_x\phi(x,t)]^2 \right\}dx$ and
$\mathcal{L}^{\text{JJ}}_j=\frac{C_J}{2}\delta\dot{\phi}_j^2+\frac{\varphi_0^2}{L_J}\cos\left(\frac{\delta\phi_j}{\varphi_0}\right)$
are the Lagrangians of the coplanar waveguide (CPW) pieces between JJs and of the JJs themselves. $\Delta=L/(N+1)$ is the spacing between JJs and $\varphi_0=\hbar/(2e)$ the rescaled quantum of flux. The JJs introduce a drop $\delta\phi_j = \phi|_{x\nearrow j\Delta} - \phi|_{x\searrow j\Delta}$ between the limits of the flux function approaching the JJ from the left, $\phi|_{x\nearrow j\Delta}$, and from the right, $\phi|_{x\searrow j\Delta}$.

\paragraph{Eigenmodes\label{eigenmodes}}
 To elucidate the underlying physics of the described CPWR we may strive after decomposing the linear part of the Lagrangian (\ref{eq:;lagrangian}) into eigenmodes which then couple via the nonlinearity of the JJs. From the Euler-Lagrange equations we get
wave equations $\partial^2_t\phi-v^2\partial^2_x\phi=0$ for the pieces of CPW between adjacent JJs. Together with the boundary conditions of vanishing current at the ends of the CPWR,
$\partial_x\phi|_{x=0}=\partial_x\phi|_{x=L}=0$, current conservation at the JJs,
$\partial_x\phi|_{x\nearrow j\Delta} = \partial_{x} \phi|_{x\searrow j\Delta}$,
and the linearized current-flux relations of the JJs, 
$-\partial_x\phi|_{x=j\Delta}/l=C_J\delta\ddot{\phi}_{j}+\delta\phi_{j}/L_J$,
these lead to a well defined eigenvalue problem.
We thus decompose $\phi(x,t) = \sum_{i} g_{i}(t) f_{i}(x)$, where 
we can sort the eigenmodes into $N$ different manifolds owing to the symmetry of the device. Each of their eigenfrequencies $\omega_{i}$ is a solution to one of the $N$ transcendental equations,
\begin{equation}\label{eigEn}
\frac{\cos\left(\frac{\omega}{v}\Delta\right)-\cos\left(n\frac{\pi}{N+1}\right)}{\sin\left(\frac{\omega}{v}\Delta\right)}=\frac{cv}{2 C_J}\frac{\omega}{\omega_p^2-\omega^2}\,,
\end{equation}
where $n\in [1,N]$. 
The eigenmode functions read
$f_i(x)=c_{i,m} \cos(\frac{\omega_i}{v}(x\bmod{\Delta})) +s_{i,m} \sin(\frac{\omega_i}{v}(x\bmod{\Delta}))$
for $x \in (m \Delta, (m+1) \Delta)$, where $i= (n,k)$ indexes all eigenmodes.
The derivation of the Eqs. (\ref{eigEn}) and the eigenmodes $f_{i}$ together with explicit expressions for the coefficients $c_{i,m}$ and $s_{i,m}$ can be found in the supplemental material~\cite{supplement}.
Note that there are also eigenmodes of the bare CPWR with $\omega_{m} = (\pi v / L) (N+1) m$ ($m \in \mathbb{N}_0$) that have current nodes at the JJs and hence do not couple to them.
\paragraph{Spectrum}
 \begin{figure}
\includegraphics[width=\columnwidth]{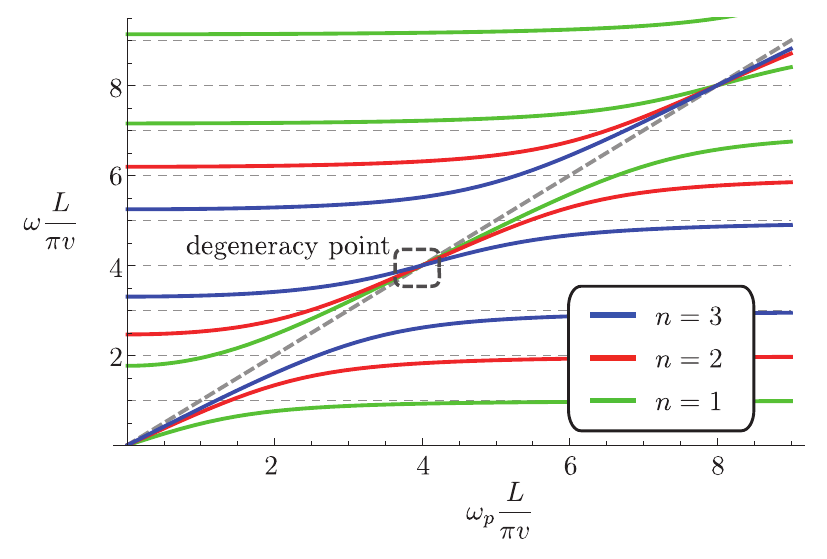}
\caption{Spectrum of CPWR with $v=0.98\times10^{8}$m/s and $Z_{0} = 50\Omega$, interrupted by 3 identical and uniformly distributed JJs with $C_{J}=1$pF and $L = 28$mm, plotted over their plasma-frequency $\omega_p$ in units of the fundamental mode frequency $(\pi v)/L$ \label{spec}. Eigenmodes of the bare CPWR that do not couple to the JJs are omitted. The dashed box marks the degeneracy point $\omega_{p} = \overline{\omega}$, on which we focus our investigation.}
\label{fig:spectrum}
\end{figure}
Fig. \ref{spec} displays the frequencies $\omega_{i}$ of a CPWR with $N=3$ JJs as a function of the plasma frequency $\omega_p$ in units of the fundamental mode frequency $\pi v/L$ of the bare CPWR. For each manifold, it resembles the spectrum of a JJ mode that is ultrastrongly coupled to specific free modes of the bare CPWR \cite{Leib2012}. For a large detuning between free mode frequency and plasma frequency the JJs and the CPWR oscillate independently. In turn for degenerate plasma and free mode frequencies, the eigenfrequencies of the combined device show an avoided crossing of the order of the eigenmode frequencies themselves, evidencing ultra-strong coupling.
The size of the avoided crossing scales as $\sim\sqrt{N}$ \cite{supplement}, similar to the Dicke model \cite{Dicke1954}. In this regime excitations of the device are strongly hybridized between the CPWR and the JJs, where the involved JJ mode can however not be traced back to a specific JJ but rather is a combined excitation of all JJs with a specific symmetry.

Note that our approach also shows that no superradiance quantum phase transition \cite{Nataf10,Viehmann11} can be observed in the cQED setups we consider, since none of the eigenmode frequencies vanishes for $\omega_{p}>0$.

 \paragraph{Quantization}
With the help of the above derived eigenmode functions we can decompose the linear part of the JJ doped CPWR into independent harmonic oscillators. Including the nonlinear terms and performing a Legendre transform we get the full Hamiltonian
$\mathcal{H}=\sum_i \frac{1}{2\eta_i}\pi_i^2+\frac{1}{2}\eta_i\omega_i^2g_i^2 + \mathcal{H}_{NL}$,
where $\eta_i=c\int_{0}^{L}f_i^2 dx+ C_J\sum_{j=1}^{N}\left(f_{i}|_{x\nearrow j\Delta}-f_{i}|_{x\searrow j\Delta}\right)^2$ is the effective mass of eigenmode $i$ \cite{footnote1}, $\pi_i=\eta_i \dot{g_i}$ the canonical conjugate momentum of $g_i$ and 
$\mathcal{H}_{NL} = -\frac{\varphi_0^2}{L_J}\sum_{j=1}^{N}\left[\cos\left(\frac{\delta\phi_j}{\varphi_0}\right)+ \frac{\delta\phi_j^2}{2\varphi_0^2}\right]$
the nonlinear part of the Hamiltonian.
We quantize the theory in the standard way by introducing lowering (raising) operators $a_{i} = \sqrt{\eta_i\omega_i/(2\hbar)} (\hat{g}_{i} + i \hat{\pi}_{i} /(\eta_{i}\omega_{i}))$ for the eigenmodes,
to get $\hat{\mathcal{H}}=\sum_i \hbar\omega_i a_i^{\dag}a_i + \hat{\mathcal{H}}_{NL}$ and write the flux drops as
$\delta\phi_j = \sqrt{2/(N+1)} \, \sin (p_{j}) \sum_{i} \lambda(\omega_{i})(a_{i}+a_{i}^{\dag})$,
where $p_{j} = \pi j/(N+1)$ and the zero point fluctuation amplitudes read
$\lambda(\omega)=\sqrt{\frac{\hbar}{2 C_{J} \omega}}\left[4 \zeta \xi_-^2 + \cot(\frac{\omega}{v}\Delta)
\frac{\omega}{v}\Delta \xi_-+\xi_+ \right]^{-1/2}$
with $\zeta = (l\Delta/L_{J})(\omega^{2}/\omega_{p}^{2})$ and
$\xi_{\pm}=(\omega_p^2 \pm \omega^2)/(2 \omega^{2})$.

\paragraph{Single-band approximation}
We have decomposed the linear part of the Hamiltonian into independent normal modes so that all coupling between the latter occurs via $\hat{\mathcal{H}}_{NL}$. This coupling is only relevant if the frequency difference between a pair of modes is comparable to their mutual coupling. 
We thus focus on a plasma frequency $\omega_p = \overline{\omega}$ with $\overline{\omega} = \pi v (N+1)/L$, where a set of $N$ eigenmodes with indices $i = (n,k=2)$ become degenerate, see Fig. \ref{fig:spectrum},
and concentrate our further discussions on the vicinity of this particularly interesting case.
Here, $k$ can be interpreted as a band index since a mode-function with index $k$ has $k-1$ nodes between any pair of adjacent JJs, and $n$ counts the modes within the band. 
 Any coupling to the remaining eigenmodes, i.e. other bands, is considerably smaller than their separation in frequency and can thus safely be neglected in a 'single-band approximation', see \cite{supplement}. 
One could also choose $k > 2$ but this would require a longer resonator and hence a larger chip.
We thus focus our analysis on modes with $k=2$ and the point where their frequencies $\omega_{n,2} \approx \overline{\omega}$, so that they
are described by the reduced Hamiltonian $[\hat{\mathcal{H}}]_{k=2} = \sum_{n} \hbar\omega_{n} a_{n}^{\dag}a_{n} + [\hat{\mathcal{H}}_{NL}]_{k=2}$ (we skip the index $k$ from now on: $a_{n} \equiv a_{n,2}$ and $\omega_{n} \equiv \omega_{n,2}$). 

\paragraph{Localized modes}
It is here most convenient to choose a basis of modes for which the mutual mode coupling via the nonlinearity is minimized.
At $\omega_p = \overline{\omega}$ where all modes with $k=2$ are perfectly degenerate this occurs for the
modes
$b_j =\sqrt{2/(N+1)}\sum_{n=1}^N \sin\left(j p_{n} \right) a_{n}$,
whose eigenfunctions have a large flux drop at a specific JJ and considerably smaller flux drops at all other JJs.
An illustration is presented in \cite{supplement}. 
We thus express the $[\hat{\mathcal{H}}]_{k=2}$ in terms of these modes and expand their interaction up to quartic order, see \cite{supplement}, to find
\begin{align} \label{Ham-b}
H&= \sum_{j=1}^N\left[\overline{\omega} b_j^{\dag}b_j-\frac{E_{C}}{2} b_j^{\dag}b_j^{\dag}b_jb_j\right]+\\
+&\overline{\omega} \sum_{j,l}^N u_{j,l} b_j^{\dag}b_l-E_{C}\sum_{j,l}^N g_{j,l}\left((b_j^{\dag}b_j^{\dag}b_j+b_j^{\dag})b_l+\text{H.c.}\right)\nonumber, 
\end{align}
with the single JJ charging energy $E_C=e^2/(2C_J)$, $u_{j,l}=\frac{2}{N+1}\sum_{n=1}^{N}\sin(j p_n)\sin(l p_n)[(\omega_{n}/\overline{\omega})-1]$, and 
$g_{j,l}=\frac{2}{N+1}\sum_{n=1}^{N}\sin(j p_n)\sin(l p_n)[(\lambda (\omega_{n})/\lambda_{0})-1]$, where $\lambda_{0} = \sqrt{\hbar/ 2 C_J\omega_p}$ is the zero point fluctuation amplitude of a single JJ, see \cite{supplement} for $\lim_{\omega_{p} \to \overline{\omega} } \lambda (\omega_{n})$.
The resulting Hamilton operator describes a set of mutually coupled oscillators with Kerr type nonlinearities of strength $E_{C}$ that can be substantial even on the single photon level.
Interestingly the coupling is not only formed by linear particle exchange, but also contains a non-linear, density assisted excitation exchange. Yet, due to the advantageous choice of modes, $b_{j}$, both the linear and nonlinear couplings are indeed weak, $|u_{j,l}|\ll1$ and $|g_{j,l}|\ll1$.
We thus have a rather unique situation, where a set of highly nonlinear modes form a narrow frequency band and can be efficiently driven by a single input tone.

\paragraph{Synchronized switching}
To explore the propagation of microwave photons through our device we assume that our CPWR is capacitively coupled to in- and output lines formed by half infinite CPWs. With these outlets we continuously drive the CPWR from one side with sinusoidal microwave signals of frequency $\omega_L=\omega_p - 4 E_C/\hbar$ and magnitude $\Omega$ in such a way that only the modes $b_{j}$ are excited.
To describe this process it suffices to add the driving term 
$H_{\Omega}=i\sin(\omega_L t)\sum_{j}\Omega_j(b_j-b_j^{\dag})$ to the Hamilton operator (\ref{Ham-b}),
where the driving amplitudes $\Omega_j=\Omega \sum_{n=1}^N\sin(n p_{j})/\sqrt{\eta_{n}}$ are a consequence of the charge quadratures of the $b_j$-modes at the driven end of the CPWR, see \cite{supplement}. 
Our device thus has the appealing property that the eigenmodes inherit the full nonlinearity of the JJs but are nonetheless well excitable by driving the resonator. 
We model the photon dissipation with a decay rate $\kappa$ for each mode so that the dynamics of our system is described by the master equation
$\dot{\rho}=\frac{i}{\hbar}\left[\rho,H_{\Omega}+H\right]+ \frac{\kappa}{2}\sum_{j=1}^N (2b_j\rho b_j^{\dag}-\rho b_j^{\dag}b_j-b_j^{\dag}b_j\rho)$,
where $\rho$ is the density matrix of all modes $b_{j}$.
Due to the high coordination number of the Hamiltonian $H$, c.f. Eq. (\ref{Ham-b}), where all modes mutually couple, a mean-field approach is expected to provide an accurate approximation. We thus decouple the individual modes $b_j$ according to
$b_j^{\dag}b_k \to \langle b_j^{\dag}\rangle b_k + b_j^{\dag}\langle b_k\rangle$ and
$(b_j^{\dag}b_j^{\dag}b_j+b_j^{\dag})b_k \to (\langle b_j^{\dag}b_j^{\dag}b_j\rangle+\langle b_j^{\dag}\rangle)b_k+(b_j^{\dag}b_j^{\dag}b_j+b_j^{\dag})\langle b_k\rangle$,
and solve the coupled equations of motion for all modes $b_j$ iteratively \cite{supplement}.

\begin{figure}
\includegraphics[width=\columnwidth]{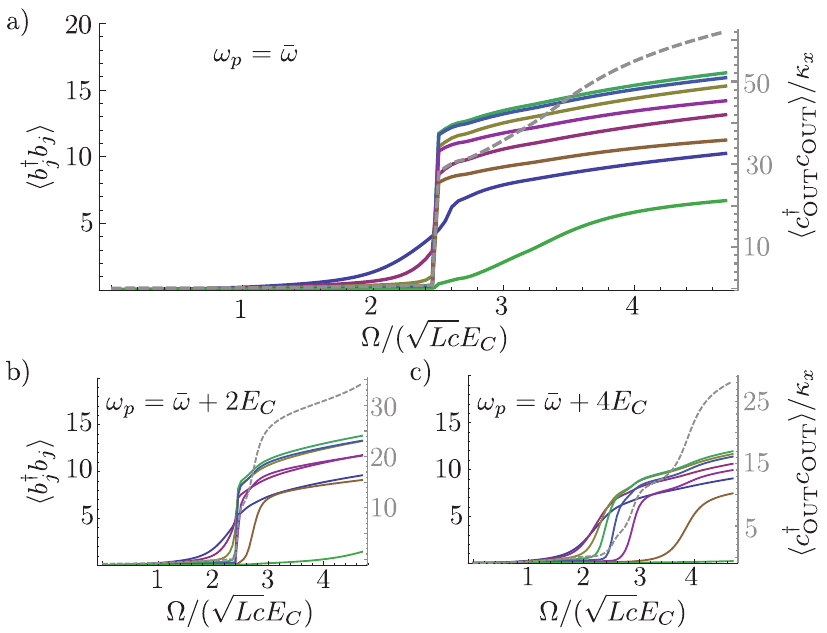}
\caption{Occupancies for the local modes $b_j$ (colored solid lines) plotted as a function of the drive strength $\Omega$ (scale at left vertical axes) at the degeneracy point $\omega_{p} = \overline{\omega}$ (a) and for plasma frequencies $\omega_p$ slightly detuned from the degeneracy point (b) and (c). The gray dashed line shows the output intensities (scale at right vertical axes). Parameters: $v=0.98\times10^{8}$m/s, $Z_{0} = 50\Omega$, $C_{J}=1$pF, and $N = 8$.}
\label{syncPlot}
\end{figure}

We find that, upon increasing the intensity of a red detuned drive for $\omega_{p} = \overline{\omega}$, all modes switch synchronously and instantly from low occupancies ($\omega_{p} < \overline{\omega}$) to high photon numbers ($\omega_{p} > \overline{\omega}$), see Fig. \ref{syncPlot}. This phenomenon can be understood as follows.
Each mode $b_{j}$ features a negative Kerr nonlinearity.
Ignoring the inter-mode couplings, one would thus expect that the combination of slightly red detuned driving and negative nonlinearity leads to a switching behavior as a function of the drive strength since the Kerr nonlinearity can be considered as an intensity dependent frequency shift \cite{Drummond1980}. Hence, upon driving the oscillator increasingly strong the frequency will drop and, for a critical driving strength, eventually come into resonance with the drive, causing a growth of oscillator excitations. Due to the different driving amplitudes $\Omega_j$ one would expect a different critical drive strength for each mode $j$. Yet, the coupling $g_{j,l}$ between the modes is such that only modes $b_j$ that are in phase amplify each other. This causes the switching of all modes to be synchronized and very sharp for the following reason.
While switching in the higher excited state the modes also undergo a quantum to classical transition. For a red detuned drive of small amplitude they behave like qubits with a $\pi$-phase delay with respect to the drive phase. After the switching into the higher excited state however the modes are in phase with the drive like harmonic oscillators with red detuned driving. As the mode with the lowest critical drive strength tries to switch it is getting weighed down because of the phase synchronizing features of the coupling $g_{j,l}$. Yet, if eventually a majority of modes switches into the higher excited state they drag the remaining modes with them, causing a very sharp and synchronized transition. If we detune the plasma frequency of the JJs from the point of degeneracy, we introduce additional mixing between the modes $b_j$, and, for detunings $\Delta\omega>E_C$, finally destroy the symmetry of the coupling that promotes synchronization of phases. As a consequence the synchronization of the switching behavior deteriorates and is eventually lost.
To determine the measurable signal in an experiment, we derive input-output relations for our CPWR,
$c_{\text{OUT}} = \sqrt{\kappa_x}\sum_{j} \tau_{j} b_j - c_{\text{IN}}$
where $\kappa_{x}$ is the decay rate into the output line and $\tau_{j} = \sqrt{2/[(N+1)\sum_n \eta_n^{-1}]} \sum_{n} (-1)^n\sin(j p_n) / \sqrt{\eta_n}$ 
and, assuming vacuum noise outside the CPWR, compute the output intensity $\langle c_{\text{OUT}}^{\dagger} c_{\text{OUT}} \rangle$.
The result is shown as gray dashed lines in Fig. \ref{syncPlot} and shows a sharp jump of $\langle c_{\text{OUT}}^{\dagger} c_{\text{OUT}} \rangle$ at the critical driving strength.

\paragraph{Experimental requirements}
There are no challenging requirements for experimentally observing the phenomena we explore here.
Photon frequencies of $6-9$ GHz as typically used in cQED setups correspond to wavelength around 14mm.
At the degeneracy point half a wave length needs to fit in between every pair of JJs which implies an overall CPWR  length of $(N+1) 0.007\text{m}$. We have chosen a phenomenological decay rate of $\kappa/(2\pi)=20$MHz for the modes $b_j$, however the synchronization is very robust with respect to dissipation and permits stronger losses.
For more details see \cite{supplement}.

\begin{acknowledgments}
We acknowledge fruitful discussions with Franco Nori. This work is part of the collaborative research centre SFB 631 and the Emmy Noether project HA 5593/1-1, both funded by the German Research Foundation (DFG). M.L. further acknowledges hospitality and support from the Center of Emergent Matter Science at RIKEN in Saitama, Japan.
\end{acknowledgments}

\bibliography{AdvNonlinRes}

\begin{thebibliography}{32}
\bibitem [{\citenamefont {for convenience with the employed transfer matrix
  technique we normalized the eigenmode functions such~that
  $f_i|_{x=0}=1$}()}]{footnote1}%
  \BibitemOpen
  \bibfield  {author} {\bibinfo {author} {\bibnamefont {for convenience with
  the employed transfer matrix technique we normalized the eigenmode functions
  such~that $f_i|_{x=0}=1$}},\ }\href@noop {} {\ }\BibitemShut {NoStop}%
\end{thebibliography}

\newpage
$ $
\newpage

\begin{widetext}
\section{Supplemental material}

\subsection{Diagonalization of the linearized Lagrangian}
Here we derive the  $N$ transcendental equations for the eigenmode frequencies, c.f. Eq. (2) of the main text with the help of a transfer matrix technique \cite{Zueco2012}. We then solve the transcendental equations to find the eigenfrequencies $\omega_{i}$ and spatial eigenmodes $f_{i}$.

The flux function $\phi_{j}$ of each CPW slice $j$, which stretches from $x=(j-1)\Delta$ to $j\Delta$, fulfills the wave equation,
\begin{equation*}
\partial^2_t\phi_{j}-v^2\partial_x^2\phi_{j}=0\,,
\end{equation*}
with phase velocity $v=1/\sqrt{lc}$. An ansatz $\phi_{j}(x,t)=g_{j}(t)f_{j}(x)$ with separation of the variables time $t$ and space $x$ leads us to the linear dispersion relation $\omega=vk$, with $k$ the wavevector, and the fundamental solution for the spatial function of the flux $f_j(x)=a_j(x)+b_j(x)$, where $a_j(x)=A_j e^{ikx}$ and $b_j(x)= B_j e^{-ikx}$.
Because of the linearization of the Josephson current-flux relations we can also separate the spatial and temporal functions in the equations of motion for the flux drops at the JJs. These linearized current-flux relations for the JJs,
\begin{equation*}
-\partial_x\phi_{j}(\Delta)=-\partial_x\phi_{j+1}(0)=\frac{l}{L_J}\left(1-\frac{\omega^2}{\omega_p^2}\right)\delta\phi_j\,,
\end{equation*} 
 can be recast in identical 2 by 2 matrices $S$ that relate the positive and negative frequency components $a_j(x)$ and $b_j(x)$ directly to the left of the JJ to their counterparts directly to the right of the JJ \cite{1367-2630-14-7-075024},
\begin{equation}\label{eq:alphadef}
\left(
\begin{array}{c}
a_{j+1}(0)\\
b_{j+1}(0)
\end{array}
\right)
=\underbrace{\left(\begin{array}{cc}
1+i\alpha & -i\alpha\\
i\alpha & 1-i\alpha
\end{array}
\right)}_{S}
\left(
\begin{array}{c}
a_j(\Delta) \\
b_j(\Delta)
\end{array}
\right)\,,
\end{equation} 
where $1/\alpha=2l/(k L_J)(1-\omega^2/\omega_p^2)$ . We moreover define the translational matrix $T(\delta x)$, which propagates the flux function along the CPW for a length $\delta x$,
\begin{equation*}
\left(
\begin{array}{c}
a_j(x+\delta x) \\
b_j(x+\delta x)
\end{array}
\right)
=\underbrace{\left(\begin{array}{cc}
e^{ik\delta x} & 0\\
0 & e^{-ik\delta x}
\end{array}
\right)}_{T(\delta x)}
\left(
\begin{array}{c}
a_j(x) \\
b_j(x)
\end{array}
\right)\,.
\end{equation*} 
We denote $D=T(\Delta)$ and relate the wavefunction parameters $a_1(0)$ and $b_1(0)$ at the left end of the CPWR to the wavefunction parameters $a_{N+1}(\Delta)$ and $b_{N+1}(\Delta)$ at the right end of the CPWR,
\begin{equation}\label{eq:PropThrough}
\left(
\begin{array}{c}
a_{N+1}(\Delta)\\
b_{N+1}(\Delta)
\end{array}
\right)
=D(SD)^N\left(
\begin{array}{c}
a_1(0) \\
b_1(0)
\end{array}
\right)\,.
\end{equation}
The current has to vanish at the ends of the CPWR, so that $\frac{1}{l}\partial_x\phi_1|_{x=0}=0$ and $\frac{1}{l}\partial_x\phi_{N+1}|_{x=\Delta}=0$ which imposes the following conditions on the spatial flux function parameters,
\begin{eqnarray*}
a_{1}(0)-b_{1}(0)&=&0\, \\
a_{N+1}(\Delta)-b_{N+1}(\Delta)&=&0\, .
\end{eqnarray*} 
With the help of these conditions, Eq. (\ref{eq:PropThrough}) can be rewritten as a homogeneous linear equation for the wavefunction parameters $a_{1}(0)$ and $b_{1}(0)$,
\begin{equation} \label{eq:homlineq}
\mathcal{X}
\left(\begin{array}{c}
a_{1}(0)\\ b_{1}(0) 
\end{array}
\right)=
\left(\begin{array}{c}
0\\0
\end{array}\right)\,,
\end{equation}
where the matrix $\mathcal{X}$ reads
\begin{equation*}
\mathcal{X} = \left(
\begin{array}{c}
1 \quad -1\\
(1\quad -1)\cdot D(SD)^N
\end{array}
\right) 
\end{equation*}
Equation (\ref{eq:homlineq}) only has nontrivial solutions if the determinant of the coefficient matrix $\mathcal{X}$ vanishes.
We thus use the condition $det(\mathcal{X}) = 0$ to determine the eigenfrequencies of the CPWR. To this end we rewrite 
\begin{equation*}
D(SD)^N=D^{\frac{1}{2}}(D^{\frac{1}{2}}SD^{\frac{1}{2}})^ND^{\frac{1}{2}}\,,
\end{equation*}
and express 
\begin{eqnarray}
D^{\frac{1}{2}}SD^{\frac{1}{2}}&=&\underbrace{(\cos(k\Delta)-\alpha\sin(k\Delta))}_{v_0}\mathds{1}+\underbrace{i(\alpha\cos(k\Delta)+\sin(k\Delta))}_{v_z}\sigma_z\underbrace{-\alpha}_{v_y} \sigma_y\nonumber\\
&=&v_0\mathds{1}+\vec{v}\cdot\vec{\sigma}\label{def:v0}\,. \label{eq:v0def}
\end{eqnarray}
in terms of Pauli matrices  $\sigma_x$,$\sigma_y$,$\sigma_z$ and the 2 by 2 identity matrix $\mathds{1}$,
Using the algebraic properties of the Pauli matrices we can further calculate arbitrary powers of the above matrix,
\begin{equation*}
(D^{\frac{1}{2}}SD^{\frac{1}{2}})^N=\underbrace{\left[\sum\limits_{n=0}^{[N/2]}\binom{N}{2n}v_0^{N-2n}(\vec{v}.\vec{v})^{n}\right]}_{\text{co}(v)}\mathds{1}+\underbrace{\left[\sum\limits_{n=0}^{[N/2]}\binom{N}{2n+1}v_0^{N-1-2n}(\vec{v}.\vec{v})^{n}\right]}_{si(v)}\vec{v}\cdot\vec{\sigma}\,,
\end{equation*}
where $[N/2]$ denotes the integer part of $N/2$.
From these expressions we find $det(\mathcal{X}) = \sin(k\Delta)\left[co(v)+si(v)v_0\right]$, which leads us to the following transcendental equation for the eigenfrequencies of the CPWR,
\begin{equation*}
\sin(k\Delta)\left[co(v)+si(v)v_0\right]=0\,.
\end{equation*}
Here, $\sin(k\Delta)=0$ singles out the eigenmodes of the bare CPWR that do not couple at all to the JJs because the current for these specific modes vanishes at the positions of the JJs. The remaining factor $\left[co(v)+si(v)v_0\right]$ can be simplified to find the transcendental equation for all other eigenfrequencies,
\begin{equation}
\sum\limits_{n=0}^{[N/2]}\binom{N+1}{2n+1}v_0^{N-2n}(v_0^2-1)^n=0\,.\label{eq:v0}
\end{equation}
As the left hand side of this equation is a polynomial of degree $N$ in the variable $v_{0}$, the equation has $N$ solutions. It is straight forward to verify that these $N$ solutions read 
\begin{equation}\label{eq:v0sol}
v_{0,n} = \cos\left(\frac{n \pi}{N+1}\right)\, .
\end{equation}
We found the solutions $v_{0,n}$ via the following physically intuitive consideration.
Taking the definition of $v_{0}$ in Eq. (\ref{eq:v0def}) and inserting the expression for $\alpha$, c.f. Eq. (\ref{eq:alphadef}), we find the equation,
\begin{equation}\label{eq:trans1}
\frac{\cos\left(\frac{\omega}{v}\Delta\right)-v_{0}}{\sin\left(\frac{\omega}{v}\Delta\right)}=\frac{cv}{2 C_J}\frac{\omega}{\omega_p^2-\omega^2}\,,
\end{equation}
Since all coefficients in the polynomial on the left hand side of Eq. (\ref{eq:v0}) only depend on the number of JJs $N$ but not on any other properties of the system, we conclude that $v_{0}$ can not depend on the system properties either (except for $N$).
Moreover, physically one expects that an eigenmode of the bare resonator is not affected by the JJs in the limit where its frequency $(\pi v/L) n$ ($n\in \mathbb{N}$) is infinitely far detuned from the plasma frequency $\omega_{p}$ of the JJs. Therefore every eigenfrequency $\omega_{i}$ of the combined system should converge to an eigenfrequency of the bare resonator in the limit where the latter is far detuned from $\omega_{p}$. In this limit, the right hand side of Eq. (\ref{eq:trans1}) vanishes. Using $\Delta = L/(N+1)$ we thus get for $\omega \to (\pi v/L) n$ the solution $v_{0,n}$, as given in Eq. (\ref{eq:v0sol}). 

Even though we found these solutions $v_{0,n}$ in the limit where the corresponding eigenfrequencies are infinitely far detuned from $\omega_{p}$ they solve Eq. (\ref{eq:v0}) for all possible system parameters because $v_{0}$ can not depend on the latter for the reasons discussed above. 
Hence inserting the solutions $v_{n,0}$ as given in Eq. (\ref{eq:v0sol}) into Eq. (\ref{eq:trans1}) we find N transcendental equations for the eigenfrequencies,
\begin{equation}\label{eigEn-sup}
\frac{\cos\left(\frac{\omega}{v}\Delta\right)-\cos\left(n\frac{\pi}{N+1}\right)}{\sin\left(\frac{\omega}{v}\Delta\right)}=\frac{cv}{2 C_J}\frac{\omega}{\omega_p^2-\omega^2}\,,
\end{equation}
where $n\in [1,N]$. The infinite number of solutions for each of these transcendental equations constitute a manifold with the same index $n$.

We now label all eigenfrequencies $\omega_{i}$ of our device with an index $i=(n,k)$, where $n$ denotes the manifold and $k$ counts the solutions within one manifold.  The associated spatial flux functions $f_i$ can be determined by propagating the initial $a_{1,i}(0)$ and $b_{1,i}(0)$ from the left border of the CPWR to position $x$ with the help of the transfer matrices $T$,$D$ and $S$. All pairs of $a_{1,i}(0)$ and $b_{1,i}(0)$ are equal, because of current conservation at the end of the CPWR and we choose them to be $a_{1,i}(0)=b_{1,i}(0)=\frac{1}{2}$ without loss of generality. This provides us with a set of orthogonal yet not normalized spatial eigenfunctions. Therefore the spatial flux eigenfunctions $f_i$ may be written,
\begin{equation*}
f_{i}(x)=\frac{1}{2}(\begin{array}{cc}1&1\end{array})T(x\bmod{\Delta})(SD)^m\left(\begin{array}{c}1\\1\end{array}\right)\,,
\end{equation*}
where $m$ is the number of JJs before the considered point in the CPWR $x$. Note that for the matrices $T$, $D$ and $S$ we have suppressed an index $i$ for readability that originates from their dependence on $k$ or $\omega$. 
Using again the algebraic properties of the Pauli matrices we get the simplified expressions for the eigenmode functions,
\begin{eqnarray}\label{eigFun}
f_i(x)&=&c_{i,m} \cos\left(\frac{\omega_{i}}{v}(x\bmod{\Delta})\right)+s_{i,m} \sin\left(\frac{\omega_{i}}{v}(x\bmod{\Delta})\right)\,,
\end{eqnarray}
with,
\begin{eqnarray*}
c_{i,m}&=&\cos\left(\pi \frac{mn}{N+1}\right)-\frac{\sin\left(\pi\frac{mn}{N+1}\right)}{\sin\left(\pi\frac{n}{N+1}\right)}\left(\cos\left(\frac{\omega_{n,k}}{v}\Delta\right)-\cos\left(\pi\frac{n}{N+1}\right)\right)\\
s_{i,m}&=&-\frac{\sin\left(\pi\frac{mn}{N+1}\right)}{\sin\left(\pi\frac{n}{N+1}\right)}\sin\left(\frac{\omega_{n,k}}{v}\Delta\right)\,.
\end{eqnarray*}

\subsection{Estimate of coupling between JJs and CPWR}
The interaction between the JJs and the bare CPWR has been modeled exactly with the microscopic approach that we chose. An alternative approach would be to diagonalize the bare CPWR and the JJs separately  and then treat the interaction between them in a perturbative manner.
The effective coupling frequency $g$ between the JJs and the bare CPWR, is given by half of the frequency difference at the avoided crossing where a mode of the bare CPWR and the JJs become degenerate, i.e. for $\omega_p=(\pi v/L) n$ for all integers $n$ that are not multiples of $N+1$. We restrict ourselves to the avoided crossings which are closest to the considered degeneracy point $\omega_{p} = \overline{\omega}$ on the lower frequency side $\omega_{p} < \overline{\omega}$. These occur for modes with $k=1$ in each manifold.
We expand the transcendental equations (\ref{eigEn-sup}) to second order in $\delta\omega = \omega - \omega_{p}$ around the degeneracy point $\omega = \omega_{p}$, and get
\begin{equation*}
\delta\omega=\pm\frac{1}{2\pi}\sqrt{\frac{Lc(N+1)}{C_J}}\quad\Rightarrow\quad g=\frac{1}{2}\sqrt{\frac{N+1}{C_J L l}}\,.
\end{equation*}
This expression has a simple physical interpretation as the coupling of two resonant oscillatory circuits where one is coupled to the current of the other has the same functional form. The coupling strength is proportional to the frequency of a fictional resonating circuit with the inductance of the one circuit and the capacitance of the other. The coupling strength $g$ grows with the square root of the number of JJs comparable to the situation in the Dicke model.

\subsection{Selection Rules for the Nonlinearity}
In this section we derive among which modes the nonlinearity can induce coupling.
The nonlinearity of the CPWR is the sum of the Josephson terms of all phase-jumps at the JJs, where quadratic terms have been subtracted. In a translation invariant chain with periodic boundary conditions the scattering of excitations due to the nonlinearity
conserves momentum which leads to a selection rule. Here we exploit the equivalent for our finite chain. As a first step to further analyze the workings of the nonlinearity we expand the polynomial of all manifold flux quadratures $\overline{\phi}_n=\sum_{k}\hat{\phi}_{n,k}=\sum_{k} \lambda(\omega_{n,k})(a_{n,k}+a_{n,k}^{\dag})$,
\begin{eqnarray}\label{eq:nonlin}
\hat{\mathcal{H}}_{NL}&=& - \frac{\varphi_0^2}{L_J}\sum_{j=1}^{N}\sum_{l=2}^{\infty}\frac{-2^l}{(2l)!(N+1)^l}\left(\sum_{n=1}^N\sin(\pi \frac{jn}{N+1})\frac{\overline{\phi}_n}{\varphi_0}\right)^{2l}\\
&=&-\frac{\varphi_0^2}{L_J}\sum_{j=1}^{N}\sum_{l=2}^{\infty}\frac{-2^l}{(2l)!(N+1)^l}\sum_{\bar{n}\in P(2l)}\prod_{m=1}^{2l}\sin(\pi \frac{j\bar{n}_m}{N+1})\frac{\overline{\phi}_{\bar{n}_m}}{\varphi_0}\nonumber \,.
\end{eqnarray}
Here $\bar{n}$ denotes a vector of length $2l$ that is an element of the set $P(2l)$ of all combinations of manifold indices $n$ of length $2l$ and $\bar{n}_{m}$ is the element number $m$ of $\bar{n}$. We want to first compute the sum over all JJs (sum over index $j$). Since the dependencies on the JJ only arise in the sine functions it suffices to examine the expression,
\begin{equation}\label{eq:phi4}
\sum_{j=1}^N \prod_{l=1}^{2l}\sin\left(\pi\frac{j \bar{n}_l}{N+1}\right)=\frac{1}{2^{2l-1}}\sum_{j=1}^{N}\sum_{\{\sigma\}}P(\sigma)\cos\left(\pi\frac{j\sum_{m=1}^{2l}\sigma_m n_m}{N+1}\right)\,,
\end{equation}
where $\{\sigma\}$ is the set of all $2^{2l-1}$ different combinations $\sigma$ of minus and plus signs, for example $\sigma_l=\{-1,1,1,1,-1,\dots,1\}$. $P(\sigma)$ is the parity of the sign combination $P(\sigma)=\prod_{m=1}^{2l}\sigma_m$, which is either $-1$ for combinations with an odd number of minus signs or $1$ for combinations with a even number of minus signs.
The sum over $j$ can for each $\nu = \sum_{m=1}^{2l}\sigma_m n_m$ be simplified with,
\begin{equation*}
\sum_{j=1}^N \cos\left(\pi\frac{j \nu}{N+1}\right)=\begin{cases} N &\mbox{if } \nu = 2(N+1)m \quad m\in\mathbb{N}_0 \\ 
-1 & \mbox{if } \nu \mbox{ even and} \quad \nu \ne 2(N+1)m \quad m\in\mathbb{N}_0 \\
0 &\mbox{if } \nu \mbox{ odd} \,. \end{cases}
\end{equation*}
To further evaluate (\ref{eq:phi4}), we thus need to analyze when $\nu$ is even and whether there are cases of $\nu = 2(N+1) m$ with $m\in\mathbb{N}_0$. Combinations of $\bar{n}_l$ that contain an odd number of odd manifold indices do not contribute because their sum can only be odd. This selection rule originates in the mirror symmetry of the device. Eigenmode functions with odd manifold index $n$ are anti-symmteric, with respect to a symmetry axis perpendicular to the CPWR through its center, while eigenmode functions with even manifold index $n$ are symmetric. Having ruled out combinations with an odd quantity of odd manifold indices, all contributing $\nu$ must be even. If there are no $\nu$ that are multiples of $2(N+1)$, then $\sum_{j=1}^N \cos\left(\pi j \nu/(N+1)\right)= -1$ for all $\nu$ and because of the equally distributed plus and minus signs of cosine terms, i.e. because $\sum_{\{\sigma\}}P(\sigma)=0$, equation (\ref{eq:phi4}) is identically zero and the specific nonlinear coupling vanishes.
There can thus only be non-vanishing couplings if there is at least one $ \nu = 2(N+1)m$ ($m\in\mathbb{N}_0$).
In fact, for every such $\nu = 2(N+1)m$ there is another even $ \nu \ne 2(N+1)m$ with opposite parity so that the two terms add up to a pre-factor $N+1$ in each non-vanishing coupling, see Eq. (\ref{eq:nonlin}). Moreover, since
\begin{equation*}
\sum_{j=1}^N \prod_{m=1}^{2l}\sin\left(\pi\frac{j \bar{n}_m}{N+1}\right)\leq \frac{N+1}{2}\,.
\end{equation*}
each coupling term between modes scales as $(N+1)^{\alpha}$ with $\alpha \le-1$.

\subsection{Linear Couplings Due to Normal Ordering of the Nonlinearity}

To confirm the accuracy of the single-band approximation applied at $\omega_{p} = \overline{\omega}$ in the main text, we here examine threading terms of the couplings between modes $a_{i}$.

As we diagonalized the linear part of the Hamilton operator for the CPWR with JJs, there are no direct exchange coupling terms of the type,
$a_i^{\dag}a_j+a_ia_j^{\dag}$
in $\mathcal{H}-\mathcal{H}_{NL}$ as given in equation (3) of the main text. However such terms emerge from the nonlinearity by virtue of the bosonic commutation relations if the raising and lowering operators are normal ordered. 
With the above derived selection rules for the nonlinearity we can compute these linear couplings between eigenmodes, c.f. eq. (\ref{eq:nonlin}). Fig. \ref{fig:couplePlot} shows these couplings for a CPWR with 8 JJs. The coupling of every  eigenmode with $k=2$ with every eigenmode with $k=1$ is calculated at the degeneracy point $\omega_p=\frac{\pi v}{L}(N+1)$ and plotted in a color coded matrix plot. The strongest coupling, occurring between the seventh and fifth eigenmode, is plotted as a function of the plasma frequency $\omega_p$. For the couplings between all eigenmodes with $k=2$ and $k=3$ we did the same calculations and plotted the result in Fig. \ref{fig:couplePlot}. We observe that the couplings are always three orders of magnitude smaller than the fundamental mode frequency of the CPWR. Therefore we can safely neglect all interactions between eigenmodes with different values of $k$. 
In agreement with the selection rules for the nonlinearity we find the checkerboard pattern for the coupling matrices showing that there is no coupling between a symmetric and an antisymmetric eigenmode.
\begin{figure}
\includegraphics[width=0.9\columnwidth]{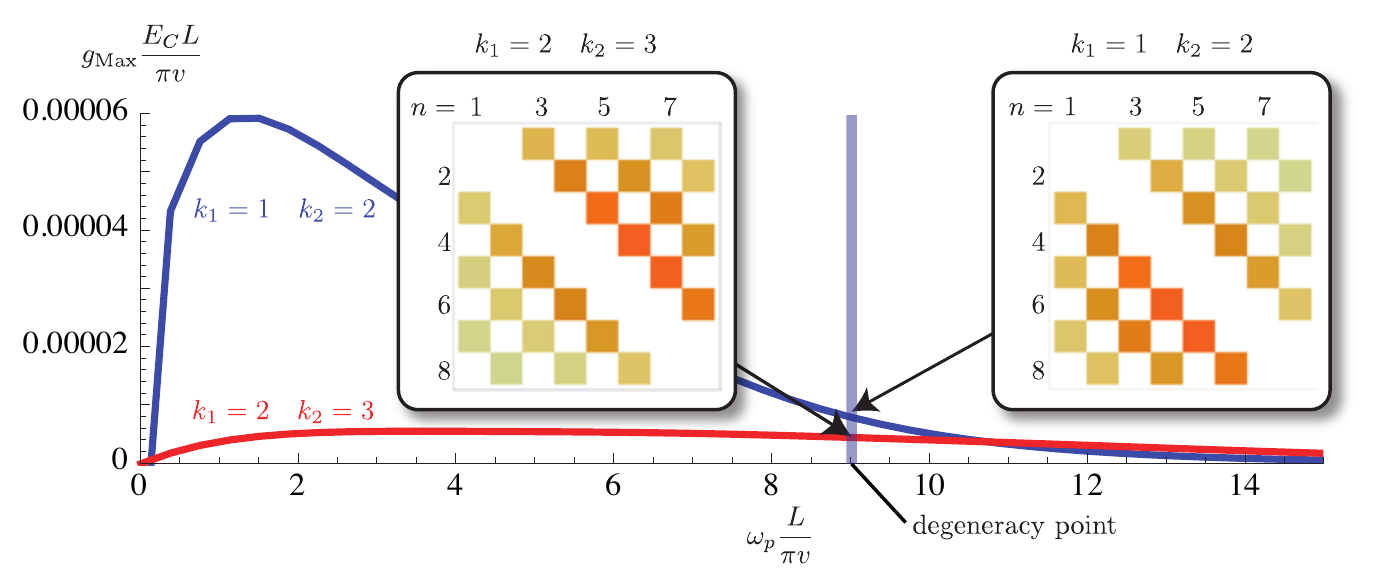}
\caption{Coupling between all eigenmodes with $k=1$ and $k=2$ and between all eigenmodes with $k=2$ and $k=3$ at the degeneracy point. For each case, the largest coupling $g_{\text{Max}}$ in units of the fundamental mode frequency $(\pi v)/L$ is plotted as a function of the plasma frequency $\omega_p$ of the JJs. }
\label{fig:couplePlot}
\end{figure}

\subsection{From Global to Local Modes} 
Here, we derive the Hamilton operator in terms of the localized modes $b_{j}$ as given in Eq. (3) of the main text.
The Hamilton operator for the eigenmodes in the vicinity of the degeneracy point is (we drop the $k$-index in the sequel),
\begin{equation*}
\left[\hat{\mathcal{H}}\right]_{k=2}=\hbar\sum_{n} \omega_{n} a_{n}^{\dag}a_{n}-\frac{\varphi_0^2}{L_J}\sum_{j=1}^{N}\sum_{l=2}^{\infty}\frac{-1^l}{(2l)!}\left(\sqrt{\frac{2}{N+1}}\sum_{n=1}^N\sin(\pi \frac{jn}{N+1})\frac{\hat{\phi}_{{n}}}{\varphi_0}\right)^{2l}\,,
\end{equation*}
where $\hat{\phi}_{{n}}= \lambda(\omega_n)(a_{{n}}+a_{{n}}^{\dag})$. Using the transformation,
\begin{equation*}
b_j =\sqrt{\frac{2}{N+1}}\sum_{n=1}^N \sin\left(\frac{\pi j n}{N+1}\right) a_{n},
\end{equation*}
we may rewrite the Hamilton operator for the degenerate eigenmodes in terms of these local Josephson junction modes,
\begin{equation} \label{eq:supplham5}
\left[\hat{\mathcal{H}}\right]_{k=2}=\sum_{j=1}^N\left[\overline{\omega}b_j^{\dag}b_j+\left(\overline{\omega}\sum_{l=1}^N u_{j,l} b_j^{\dag}b_l\right) -\frac{\varphi_0^2}{L_J}\sum_{m=2}^{\infty}\frac{(-1)^m}{(2m)!}\left(\frac{\hbar}{2\omega_p C_J\varphi_0^2}\right)^m\left((b_j+b_j^{\dag})+\sum_{l=1}^Ng_{j,l}(b_l+b_l^{\dag})\right)^{2m}\right]\,.
\end{equation}
with,
\begin{align*}
u_{j,l}&=\frac{2}{N+1}\sum_{n=1}^{N}\sin\left(\frac{j n \pi}{N+1} \right)\sin\left(\frac{l n \pi}{N+1} \right)\left[\frac{\omega_{n}}{\overline{\omega}}-1\right]\quad \text{and}\\
g_{j,l}&=\frac{2}{N+1}\sum_{n=1}^N\sin\left(\frac{j n \pi}{N+1} \right)\sin\left(\frac{l n \pi}{N+1} \right)\left[\frac{\lambda (\omega_{n})}{\lambda_{0}}-1\right]\,,
\end{align*}
where
\begin{equation*}
\lambda_{0} = \sqrt{\frac{\hbar}{2 C_{J} \omega_p}}
\end{equation*}
are the zero point flux fluctuations of a single JJ and
\begin{equation*}
\lim_{\omega_p\to\overline{\omega}}\lambda(\omega_{n})=\sqrt{\frac{\hbar}{2 C_{J} \omega_p \left(1+\frac{\Delta c}{4 C_{J}} \left[1+\cos(n\frac{\pi}{N+1})\right]^{-1}\right)}}\,,
\end{equation*}
is the magnitude of the zero point flux fluctuations of a single JJ with capacitance $C_J$, renormalized by the mode dependent capacitance of the CPWR. Note that naively taking $\lim_{\omega_p\to\overline{\omega}}\lambda(\omega_{n}) = \lambda (\omega_{p})$ does not lead to correct results since $\cot(\frac{\omega}{v}\Delta)$ diverges for $\omega_{p} \to \overline{\omega}$. We thus expressed $\cot(\frac{\omega}{v}\Delta)(\omega_{p}^{2}-\omega^{2})$ with the help of the transcendental equation for the eigenfrequencies, c.f. eq.: (\ref{eigEn-sup}), to get the correct limit.
Note that if the zero point flux fluctuations $\lambda (\omega_{n})$ of the $N$ degenerate eigenmodes were all equal, all modes $b_{i}$ would decouple and the 
Hamilton operator would decompose into a sum of independent Hamilton operators describing identical JJs.
Here however we need to consider the nonuniform coupling to the microwave drive since the zero point flux fluctuations of the eigenmodes differ from mode to mode. We thus
get a coupling between the modes. For the chosen JJ capacitances $C_J$ and resonator capacitance $Lc$ this coupling is small, i.e. $\text{max}(|u_{j,l}|)\ll 1$ and $\text{max}(|g_{j,l}|)\ll 1$. For this reason we may only keep terms that couple modes up to linear order in $g_{j,l}$ in the Hamilton operator. Importantly, $\text{max}(|g_{j,l}|)\ll 1$, suppresses correlated tunneling \cite{Jin2013} as the corresponding terms would be higher than linear order in $g_{j,l}$. Keeping only terms up to quartic order in the flux field amplitudes $\hat{\phi}_n$ in Eq. (\ref{eq:supplham5}) and 
performing a rotating wave approximation we arrive at the Hamiltonian $H$ given in equation (7) of the main text.

Whereas the eigenfunctions of the modes $a_{n}$ have comparable drops at all JJs, the eigenfunctions of the modes $b_{j}$ have a large flux drop at a specific JJ and considerably smaller flux drops at all other JJs. For an illustration, eigenfunctions of both classes of modes are plotted in figure \ref{fig:fluxFun}.
 
\begin{figure}
\includegraphics[width=0.8\columnwidth]{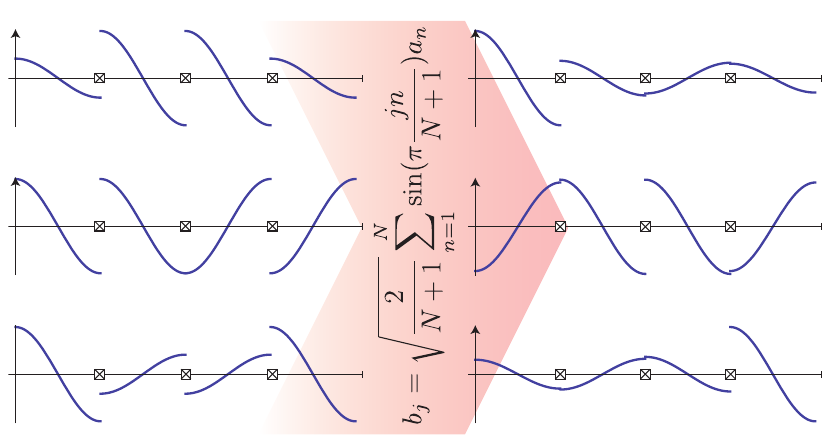}
\caption{Eigenmode functions with $k=2$ in arbitrary units at the degeneracy point for a CPWR (horizontal lines) with three JJs (crossed boxes). Left: eigenmode functions associated to the mode operators $a_n$. Right: eigenmode functions of modes $b_j$ that minimize nonlinearity induced mixing.}
\label{fig:fluxFun}
\end{figure}

\subsection{Coherent Drive and Dissipation}
We couple the CPWR capacitively to a half infinite CPW and excite the CPWR with a coherent drive tone $\phi_{ext}$. The energy stored in the coupling capacitance $C_c$ is,
\begin{equation*}
\tilde{H}_{\Omega}=\frac{C_c}{2}\left(\dot{\phi}_{ext}-\dot{\phi} |_{x=0}\right)^2=\frac{C_c}{2}\left(\dot{\phi}_{ext}^2+\dot{\phi}^2|_{x=0}\right)-C_c\dot{\phi}_{ext}\dot{\phi}|_{x=0}\,.
\end{equation*}
We consider only small coupling capacitances and neglect a small renormalization of the eigenmode frequencies caused by the first two terms in the expanded coupling energy. Moreover we choose the drive frequency to be near resonance with the modes $a_{n}$ and neglect any coupling to other modes of the CPWR. Expressing the internal flux field $\phi$ at the driven side of the resonator in terms of the eigenmodes $a_{n}$ we arrive at,
\begin{equation*}
\left[\tilde{H}_{\Omega}\right]_{k=2}\approx H_{\Omega}=-C_c \dot{\phi}_{ext}\sum_{n} \dot{g}_{n} f_{n}(x)|_{x=0}=i \dot{\phi}_{ext}\sum_{n}\sqrt{\frac{\hbar \omega_p}{2\eta_{n}}}(a_{n}-a_{n}^{\dag})\,.
\end{equation*}
Here we already introduced the lowering and raising operators of the eigenmodes of the CPWR and used the special normalization we have chosen for the eigenmode functions that they are all equal to one at the beginning of the CPWR. Next we unitarily transform to the $b_j$ modes and express the classical drive in terms of $\dot{\phi}_{ext}=\Omega \sqrt{\frac{N+1}{\hbar\omega_pC_c^2}}\sin(\omega_L t)$ to get,
\begin{equation*}
H_{\Omega}=i\sin(\omega_L t)\sum_{j=1}^N\underbrace{\Omega \left(\sum_{n=1}^N\frac{\sin\left(\pi\frac{j n}{N+1}\right)}{\sqrt{\eta_{n}}}\right)}_{\Omega_j}(b_j-b_j^{\dag})\,.
\end{equation*}
Because of varying eigenmode capacitances $\eta_{n}$ and varying values of the flux functions for modes $b_j$, at the side driven by the microwave tone, we get different effective driving strengths $\Omega_j$. A similar derivation leads to the input output relation used in the main text.

There are two different types of dissipative processes. Excitations decay through the capacitively coupled ends of the CPWR into the half infinite CPWs and there is dissipation due to two-level fluctuators in the JJs or the substrate material of the cQED setup.  The latter source of dissipation is the same for every mode $b_j$ since we assume the quality of every JJ to be the same. Yet decay through the ends of the CPWR may be different for the individual modes $b_{j}$ as they do not couple with the same strength to the in- and output CPWs. As we found this inhomogeneity to be very small, we opted to neglect it in this description.
Therefore we here include dissipative processes with a standard master equation technique,
 \begin{equation*}
 \dot{\rho}=\frac{i}{\hbar}\left[\rho,H_{\Omega}+H\right]+\frac{\kappa}{2}\sum_{j=1}^N \left(2b_j\rho b_j^{\dag}-(\rho b_j^{\dag}b_j+b_j^{\dag}b_j)\right)\,,
 \end{equation*}
where $\kappa$ is the phenomenological decay rate assumed to be equal for all modes $b_j$. Assuming that the dominant dissipation mechanism is relaxation in the JJs, we employ here an independent bath approximation for the modes $b_{j}$. 

\subsection{Mean Field Approximation in the Driven Dissipative Regime}
Due to a dynamical balance between constant injection and the leakage of microwave photons, as discussed in the preceding section, a steady state emerges.
 The effective Hamilton operator $H$ that we derived above constitutes a set of nonlinear oscillators with interactions between every oscillator. Because of this high coordination number together with weak couplings $|u_{j,l}|,|g_{j,l}| \ll 1$ a mean field approach to investigate the steady state is reasonable. We reduce the coupling to an interaction with the respective mean fields,
 \begin{align*}
 b_j^{\dag}b_k &\to \left\langle b_j^{\dag}\right \rangle b_k + b_j^{\dag}\left\langle b_k\right\rangle \quad \text{and}\\
 \left(b_j^{\dag}b_j^{\dag}b_j+b_j^{\dag}\right)b_k &\to \left(\left\langle b_j^{\dag}b_j^{\dag}b_j\right\rangle+\left\langle b_j^{\dag}\right \rangle\right)b_k + \left(b_j^{\dag}b_j^{\dag}b_j+b_j^{\dag}\right)\left\langle b_k\right\rangle\,.
 \end{align*}
 In a first step we calculate the mean fields $\left\langle b_j^{\dag}\right\rangle$ and $\left\langle b_j^{\dag}b_j^{\dag}b_j\right\rangle$ individually for every mode $b_j$ and ignore the coupling. Then we update the driving amplitudes,
 \begin{equation*}
 \Omega_j\to\Omega_j -i \overline{\omega} \sum_{k\neq j}u_{j,k} \left\langle b_j^{\dag}\right\rangle + i E_{C} \sum_{k\neq j}g_{j,k}\left(\left\langle b_j^{\dag}b_j^{\dag}b_j\right\rangle +2 \left\langle b_j^{\dag}\right\rangle\right)\,,
 \end{equation*}
 and introduce another nonlinear driving, 
 \begin{equation*}
 H_{\Omega}^{\text{nonlin}}= -E_{C} \sum_{j=1}^{N}\sum_{k\neq j} g_{j,k} \left( b_j^{\dag}b_j^{\dag}b_j \left\langle b_k\right\rangle + b_j^{\dag}b_jb_j \left\langle b_k^{\dag}\right\rangle\right)\,.
 \end{equation*}
 With the updated driving, which takes into account the driving of the modes among each other, we again calculate the mean fields $\left\langle b_j^{\dag}\right\rangle$ and $\left\langle b_j^{\dag}b_j^{\dag}b_j\right\rangle$ and iterate the procedure until it converges.
 
\subsection{Experimental Parameters}

 To examine the driven-dissipative dynamics of the CPWR at degeneracy of all its eigenmodes, as described in the main article, one has to make sure at first that the plasma frequency at the degeneracy point is within the bandwidth of the detection chain  and well below the superconducting gap. Typically one uses frequencies of $6-9$ GHz in circuit quantum electrodynamics setups. This implies half wave CPWRs of about 7 mm length at phase velocities of $v=0.98 \cdot 10^{8}\frac{\text{m}}{\text{s}}$ in CPWs. At the degeneracy point half a wave length has to fit in between neighboring JJs which implies an overall CPWR length of $(N+1) 0.007\text{m}$. Additionally the CPWR has to be in the phase regime where zero point flux fluctuations are small compared to the rescaled quantum of flux $\lambda(\omega_i)<\varphi_0$ which is why we have chosen to shunt each JJ with an additional Capacitance $C_J=1$pF. To be able to observe the transition of synchronized to non-synchronized JJ-modes, one has to change the magnitude of the Josephson inductance. This can either be achieved by designing different setups with different sizes of JJs or by using dc-superconducting quantum interference devices whose effective Josephson inductance can be tuned by threading a flux bias through their loops. The degeneracy point can be reached, given the above set of parameters, at $L_J=2.9$nH. Therefore the Josephson inductance has to be tunable around $L_J=2.9$nH. For our calculations in the driven-dissipative regime we have chosen a phenomenological decay rate of $\kappa/(2\pi)=20$MHz for the modes $b_j$. The synchronization effect however is very robust against dissipation and the decay rates of actual experimental setups may be larger without any impact on synchronization. The suitable choice of parameters is summarized in table \ref{paramtable}.
 \begin{table}
 \begin{ruledtabular}
     \begin{tabular}{ l|l  l|l }
     	\multicolumn{2}{l}{Resonator} & \multicolumn{2}{l}{Josephson Junctions}\\ \hline
 	Length&$L=(N+1)0.007 \text{m}$ & Shunting Capacitance  & $C_J= 1\text{pF}$\\
 	Wave Impedance& $Z_0=50\Omega$ & Josephson Inductance & tunable $L_J\approx 2.9 \text{nH}$ \\
 	Phase Velocity & $v=0.98 \cdot 10^{8}\frac{\text{m}}{\text{s}}$& $b_j$ decay rate & $\frac{\kappa}{2\pi}=20$MHz
     \end{tabular}
\end{ruledtabular}
\caption{Parameters for an experimental realization.}
\label{paramtable}
 \end{table}

\end{widetext}
\end{document}